\documentstyle[12pt]{article}
\newcommand{\scs}{\setcounter{equation}{0}}
\renewcommand{\theequation}{\thesection.\arabic{equation}}
\def\bal{$\alpha$\kern-.7em\hbox{$\alpha$}}
\def\bxi{$\xi$\kern-.5em\hbox{$\xi$}}

\begin{document}
\begin{flushright}
EPHOU-96-007\\
August, 1996 \\
\end{flushright}
\vspace{10mm}

\centerline{\Large
FLUX STATE IN VON NEUMANN LATTICE AND}
\vspace{5mm}
\centerline{\Large 
FRACTIONAL HALL EFFECT}
\vspace{5mm}
\centerline{\Large
Kenzo {\sc Ishikawa} and Nobuki {\sc Maeda}}
\vspace{10mm}
\centerline{\it 
Department of Physics, Hokkaido University, Sapporo 060}
\vspace{15mm}
\centerline{\Large abstract}

Formulation of quantum Hall dynamics using von Neumann lattice 
of guiding center coordinates is presented. 
A topological invariant expression of the Hall conductance is given 
and a new  mean field theory of the fractional Hall effect 
based on flux condensation is proposed. 
Because our mean field Hamiltonian has the same form as Hofstadter 
Hamiltonian, 
it is possible to understand characteristic features of the 
fractional Hall effect from Hofstadter's  spectrum. 
Energy gap and other physical quantities are computed and 
are compared with the experiments. 
A reasonable agreement is obtained. 
\newpage
\baselineskip=24pt

\section{Introduction}\scs

Systems of two-dimensional electrons in a perpendicular strong magnetic field 
are giving many exciting physics these days\cite{r1}. 
Electrons in these systems have discrete energies with finite maximum 
degeneracy\cite{r2}. 
Among infinitely many representations, an invariant representation under two 
dimensional translation is convenient for many purposes. 
Invariant representation 
under continuous translations does not exist due to the magnetic 
field but lattice translational invariant one does exist. 
von Neumann lattice coherent state representation\cite{r3}is such 
representation and we have used it to verify the integer quantum Hall 
effect. 
We give a new representation of von Neumann lattice in this paper, and  
propose a new mean field theory of the quantum Hall liquid 
based on flux condensed state. 
The fractional Hall state\cite{r4} is regarded as a kind of integer Hall 
effect due to the condensed flux, and has a large energy gap in our theory. 

It is convenient to use relative coordinates\cite{r5} which 
are perpendicular 
to velocity operators and guiding center coordinates for describing 
two-dimensional electrons in the perpendicular magnetic field. 
Coherent states in guiding center coordinates have minimum spatial 
extensions and its suitable complete subset whose element has 
discrete eigenvalues defined on lattice site are used. 
Field operators in the present representation have two-dimensional lattice 
coordinates in 
addition to Landau level index, and the many-body theory is described by a    
lattice field theory of having internal freedom. 
Exact identities such as current conservation and Ward-Takahashi 
identity\cite{r3} derived from current conservation are written in a 
transparent way and play the important roles for establishing 
an exact low-energy theorem of the 
quantum Hall effect. 
We are able to expess flux state in a symmetric way, also. 

One of the hardest and most important thing of the fractional Hall 
effect is to find a mechanism of generating energy gap. 
Without interactions there is no 
energy gap at any fractional filling state. Hence interaction generates gap. 
Once the gap is generated, fluctuations are suppressed. 
Higher order corrections 
are expected to be small if starting approximate ground state has 
the energy gap and close to eigenstate. 

Flux state has a modified symmetry under translation and can have energy 
gap at certain filling states. 
We study flux states of the fractional Hall system.

In our mean field theory of the fractional quantum Hall effect, 
an order parameter is a magnitude of the flux condensation. 
Because the system 
changes its property drastically with the change of flux it is quite natural 
to treat the flux as an order parameter. Due to the dynamical fux, our mean 
field Hamiltonian is close to that 
of Hofstadter\cite{r6} and new integer Hall effect occurs within the lowest 
Landau level space of the external magnetic field. 
$ $From Hofstadter's analysis and others\cite{r7} ground state has 
the lowest energy and the largest energy gap, 
if its flux is proportional to filling factor. 
We regard, in fact, these states as the fractional Hall states. 
Especially the principal series at $\nu=p/(2p\pm1)$ 
satisfy the self-consistency condition and have large energy gap. 
They can be observed even in the Hofstadter's spectrum of butterfly shape. 

Our mean field theory has similarities also with a mean field theory of 
composite fermion\cite{r8} in regarding the fractional 
Hall effect as a kind of integer Hall effect. 
However, ours includes interaction effects in the space of the lowest 
Landau level at the mean field level and so the effective mass, 
energy gap, and other quantities of the fractional quantum Hall states 
are close to the observed value even in the lowest order. 
These features are seen in ours but are not seen in the 
composite fermion mean field theory. 
We, also, found our ground
state energy has slightly higher energy than that of Laughlin for 
$\nu=1/3$ case. 

Due to the energy gap, fluctuations are small in the fractional quantum Hall 
state. 
Higher order corrections are small  and the perturbative expansions        
converge well, generally. 
Exact half-filling is, however, exceptional and there is no energy gap. 
Fermi surface is composed of isolated points in the lowest order, but the 
fluctuations are extremely large. 
So, the structure around the Fermi energy may be changed drastically 
from that of the lowest order by interactions at the half-filling. 
Thus our mean field may not be good at the exact half-filling. 

The paper is organized in the following way. 
In section 2, we formulate von Neumann lattice representation and 
verify the integer Hall effect. 
In section 3 flux state mean field theory on the von Neumann lattice is 
formulated and is compared with the existing experiments in Section 4. 
Conclusions are given in section 5.

\section{Quantum Hall dynamics on von Neumann lattice}\scs

Quantum Hall system is described by the following Hamiltonian, 
\begin{eqnarray}
&H=\int d{\bf x}[\Psi^\dagger(x){({\bf p}+e{\bf A})^2\over 2m}\Psi(x)+
\rho(x){V(x-y)\over2}\rho(y)], \\
&\partial_1 A_2-\partial_2 A_1=B,\ \rho(x)=\Psi^\dagger(x)\Psi(x),
\ V(x)={e^2\over\kappa}{1\over\vert x\vert}.
\nonumber
\end{eqnarray}
We ignore the disorders in this paper. 
It is convenient to use the following two sets of variables,
\begin{eqnarray}
&\xi={1\over eB}(p_y+eA_y),\ X=x-\xi,\\
&\eta=-{1\over eB}(p_x+eA_x),\ Y=y-\eta,\nonumber
\end{eqnarray}
which satisfy, 
\begin{eqnarray}
&[\xi,\eta]=-[X,Y]=-i{\hbar\over eB},\nonumber \\ 
&[\xi,X]=[\xi,Y]=0,\\ 
&[\eta,X]=[\eta,Y]=0.\nonumber 
\end{eqnarray}
We expand the electron field operators with a complete set of base 
functions, $f_l(\xi,\eta)\otimes\vert R_{m,n}\rangle$, which are 
defined by,
\begin{eqnarray}
&{({\bf p}+{\bf A})^2\over 2m}f_l={e^2 B^2\over 2m}(\xi^2+\eta^2)f_l=
E_l f_l,\ E_l={\hbar eB\over 2m}(2l+1)\\
&(X+iY)\vert R_{m,n}\rangle=a(m+in)\vert R_{m,n}\rangle,
\ a=\sqrt{2\pi\hbar\over eB}.
\nonumber
\end{eqnarray}
The coherent states on von Neumann lattice are constructed as, 
\begin{eqnarray}
&\vert R_{m,n}\rangle=(-1)^{mn+m+n}e^{A^\dagger\sqrt{\pi}(m+in)-
A\sqrt{\pi}(m-in)}\vert 0\rangle,\\
&A=\sqrt{eB\over 2\hbar}(X+iY),\ [A,A^\dagger]=1.
\nonumber
\end{eqnarray}
The expressions of electron field are given by,
\begin{eqnarray}
&\Psi(x)=\sum a_l(m,n)f_l(\xi,\eta)\otimes\vert R_{m,n}\rangle,\\
&\Psi^\dagger(x)=\sum a^\dagger_l(m,n)
f_l(\xi,\eta)\otimes\langle R_{m,n}\vert,
\nonumber
\end{eqnarray}
and they are substituted into the action integral,
\begin{eqnarray}
S&=&\int d{\bf x}\Psi^\dagger(x)i\hbar{\partial\over\partial t}\Psi(x)
-H \\
&=&\sum\langle R_{m_1,n_1}\vert R_{m_2,n_2}\rangle a^\dagger_l(R_1)(
i\hbar{\partial\over\partial t}-E_l)a_l(R_2)+
\int d{\bf k}\rho({\bf k}){V({\bf k})\over 2}\rho(-{\bf k}).
\nonumber
\end{eqnarray}
By using the conjugate momentum to the lattice site, ${\bf R}_{m,n}$, 
the action is written as,
\begin{eqnarray}
S=&\sum\sum a^\dagger_l({\bf p}_1)e^{i\phi({\bf p},{\bf N})}
\beta^*({\bf p}_1)\beta({\bf p}_2)(
i\hbar{\partial\over\partial t}-E_l)a_l({\bf p}_2)
\delta({\bf p}_1-{\bf p}_2+{2\pi\over a}{\bf N}) \nonumber\\
&+\int d{\bf k}\rho({\bf k}){V({\bf k})\over 2}\rho(-{\bf k}),
\end{eqnarray}
\begin{eqnarray*}
\beta^*({\bf p})&=&2^{1\over4}e^{-{a^2\over 4\pi}p^2_x}
\Theta_1({ia\over 2\pi}(p_x+ip_y),i),\nonumber\\
\beta^*({\bf p}+{2\pi\over a}{\bf N})&=&e^{i\phi({\bf p},{\bf N})}
\beta^*({\bf p}),\nonumber\\
\phi({\bf p},{\bf N})&=&N_x(ap_y+\pi)+N_y\pi,\nonumber\\
&&-\pi/a\leq p_i\leq\pi/a, \nonumber
\end{eqnarray*}
where $\Theta_1(x,i)$ is the elliptic theta function of 
the first kind\cite{r9} 
and ${\bf N}$ is a two-dimensional vector which has integers as components. 
The momentum is conserved and its eigenvectors are orthogonal 
if their eigenvalues are different but they are not normalized. 
We introduce normalized operators, 
\begin{eqnarray}
b_l({\bf p})&=&\beta({\bf p})a_l({\bf p}),\\
b^\dagger_l({\bf p})&=&\beta^*({\bf p})a^\dagger_l({\bf p}).
\nonumber
\end{eqnarray}
$ $From the definition, $\beta({\bf p})$ vanishes at ${\bf p}=0$. 
This reflects a constraint of the coherent states 
on von Neumann lattice\cite{r10}, 
but it causes no difficulty in our method. 
They satisfy,
\begin{eqnarray}
&\{b_{l_1}({\bf p}_1),b^\dagger_{l_2}({\bf p}_2)\}=
\sum_{{\bf N}}e^{-i\phi({\bf p}_1,{\bf N})}\delta({\bf p}_1-{\bf p}_2
+{2\pi\over a}{\bf N}),\\
&b_l({\bf p}+{2\pi\over a}{\bf N})=e^{-i\phi({\bf p}_1,{\bf N})}b_l(
{\bf p}).\nonumber
\end{eqnarray}
The action is written as,
\begin{eqnarray}
S=\sum_l\sum_{{\bf N},{\bf p}_1,{\bf p}_2}
&b^\dagger_l({\bf p}_1)e^{i\phi({\bf p}_1,{\bf N})}(
i\hbar{\partial\over\partial t}-E_l)b_l({\bf p}_2)
\delta({\bf p}_1-{\bf p}_2+{2\pi\over a}{\bf N}) \nonumber\\
&+\int d{\bf k}\rho({\bf k}){V({\bf k})\over 2}\rho(-{\bf k}),
\end{eqnarray}
\begin{equation}
\rho({\bf k})=\sum b^\dagger_{l_1}({\bf p}_1)b_{l_2}({\bf p}_2)
\delta({\bf p}_1-{\bf p}_2-{\bf k}+{2\pi\over a}{\bf N})
(l_1\vert e^{i{\bf k}\cdot\hbox{\bxi}}\vert l_2)e^{i\phi({{\bf p}_1+
{\bf p}_2\over2},{\bf N})+i{a^2\over 4\pi}k^y(p_1^x+p_2^x)}.
\end{equation}
The current operator is also written as,
\begin{eqnarray}
&j_i({\bf k})=\sum b^\dagger_{l_1}({\bf p}_1)b_{l_2}({\bf p}_2)\delta(
{\bf p}_1-{\bf p}_2-{\bf k}+{2\pi\over a}{\bf N})
(l_1\vert v_i e^{i{\bf k}\cdot\hbox{\bxi}}\vert l_2)e^{i\phi({{\bf p}_1+
{\bf p}_2\over2},{\bf N})
+i{a^2\over 4\pi}k^y(p_1^x+p_2^x)},
\nonumber\\
&{\bf v}={eB\over m}(-\eta,\xi).
\end{eqnarray}

The commutation relations (2.10), and the action (2.11), have no 
singularity in ${\bf p}$. 
The zero of $\beta({\bf p})$ in Eq.(2.9) does not cause any problem. 
Hence meaningful theory is defined in this way contrary to naive 
expectations. 
To make sure this point further, we solve the one impurity problem 
by the present representation. We find an agreement with 
previous results. Namely eigenvectors are localized around the 
impurity position if corresponding eigenvalues are isolated and 
located between Landau levels. 
Their energies are moved with the impurity, also. 
Appendix A is devoted for the short range impurity problem. 

$ $From Eqs.(2.10) and (2.12), 
we have commutation relations between charge density and field operators, 
\begin{eqnarray}
&[\rho({\bf k}),b({\bf p})]=-(l_1\vert e^{i{\bf k}\cdot\hbox{\bxi}}
\vert l_2)
e^{i{a^2\over 4\pi}k_y(2p_x-k_x)}b_{l_2}({\bf p}-{\bf k})
e^{i\pi N_x N_y+iak_y N_x},\nonumber\\
&\vert {\bf p}-{\bf k}+{2\pi\over a}{\bf N}\vert\leq {\pi\over a},\\
&[\rho({\bf k}),b^\dagger({\bf p})]=
(l_1\vert e^{i{\bf k}\cdot\hbox{\bxi}}\vert l_2)
e^{i{a^2\over 4\pi}k_y(2p_x+k_x)}b^\dagger_{l_1}({\bf p}+{\bf k})
e^{-i\pi N_x N_y-iak_y N_x},\nonumber\\
&\vert {\bf p}+{\bf k}-{2\pi\over a}{\bf N}
\vert\leq {\pi\over a}.\nonumber
\end{eqnarray}
For the momentum $\bf p$ in the fundamental region and infinitesimal 
$\bf k$, $\bf N$ vanishes. 
The right-hand sides have linear terms in $\bf k$. 
Hence Ward-Takahashi identity between the vertex part and the propagator 
is modified from that of the naive one. 
We introduce the unitary operator $U({\bf p})$ which satisfies
\begin{equation}
\delta_{l_1,l_2}{\partial\over\partial p_i}U({\bf p})+
\{i(l_1\vert\xi_i\vert l_2)+i{a^2\over 2\pi}p_x\delta_{l_1,l_2}\}U(p)=0,
\end{equation}
and make transformation of the propagator and the vertex part as,
\begin{eqnarray}
&\tilde S({\bf p})=U({\bf p})S^{(0)}({\bf p})U^\dagger({\bf p}),\\
&\tilde \Gamma_\mu({\bf p}_1,{\bf p}_2)=U({\bf p}_1)
\Gamma_\mu({\bf p}_1,{\bf p}_2)U^\dagger({\bf p}_2).\nonumber
\end{eqnarray}
They satisfy, then, 
\begin{equation}
\tilde \Gamma_\mu(p,p)={\partial\tilde 
S^{-1}(p)\over\partial p_\mu}.
\end{equation}

The current correlation function in the momentum representation is 
written as, 
\begin{eqnarray}
\pi_{\mu\nu}(q)&=&\langle{\rm T}(j_\mu(q_1)j_\nu(q_2))\rangle \\
&=&\sum S^{(0)}_{l_1,l_4}(p_1)S^{(0)}_{l_3,l_2}(p_3)
(l_1\vert\Gamma_\mu e^{i{\bf q}\cdot\hbox{\bxi}}\vert l_2)
(l_3\vert\Gamma_\nu e^{-i{\bf q}\cdot\hbox{\bxi}}\vert l_4)
\times \nonumber \\
&&\delta({\bf p}_1-{\bf p}_3-{\bf p}_1+{2\pi\over a}{\bf N})
\delta(-{\bf q}_1-{\bf p}_2+{2\pi\over a}({\bf M}+{\bf N})),
\nonumber\\
S^{(0)}_{l_1,l_4}(p_1)&=&{1\over p_1^0-E_{l_1}}\delta_{l_1,l_4},
\nonumber
\end{eqnarray}
in the lowest order of the interaction. 
The Hall conductance is the slope of $\pi_{\mu\nu}(q_1)$ at the origin 
and is expressed, under the use of the transformed propagator and 
the Ward-Takahashi identity (2.17), as, 
\begin{equation}
\sigma_{xy}={e^2\over h}{1\over 24\pi^2}\int d^3 p\epsilon_{\mu\nu\rho}
{\rm Tr}[{\partial\tilde S^{-1}(p)\over\partial p_\mu}\tilde S(p)
{\partial\tilde S^{-1}(p)\over\partial p_\nu}\tilde S(p)
{\partial\tilde S^{-1}(p)\over\partial p_\rho}\tilde S(p)].
\end{equation}
The right-hand side is a three-dimensional winding number of the mapping 
defined by the propagator $\tilde S(p)$. 
The space of the propagator is decomposed into SU(2) subspace, 
a space spanned by $l$-th Landau level and $(l+1)$-th Landau level 
from Eq.(2.18). 
Note also that $S^{(0)}(p)$ is diagonal in $l_i$. 
In this subspace, $\tilde S(p)$ is defined on the torus and the Hall 
conductance becomes integer multiple of $e^2\over h$ in the 
quantum Hall regime where there is no two-dimensionally extended 
states around the Fermi energy. 
The value is stable under perturbation effects such as interactions, 
disorders, and others as far as the series converge, as were shown by 
Coleman, Hill\cite{r11}, and others\cite{r12}\cite{r3}. 

The value of $\sigma_{xy}$ stays constant while the Fermi energy is 
moved if there is no singularity involved. 
This occurs actually at the quantum Hall regime where there is 
no two-dimensionally extended states but there are only 
localized states with discrete energies or one-dimensionally extended 
states. 
In this situation, the value is computed by the lowest order calculation 
and has no correction from the higher order corrections.

\section{Flux state mean field theory}\scs

We propose a new mean field theory based on flux state on von Neumann 
lattice in this section. 

The dynamical flux is generated by interactions 
and plays the important role in our mean field theory. 
It is described by a lattice Hamiltonian, which is due to 
the external magnetic field, and by the  dynamical magnetic flux due to 
interaction, although the original electrons are defined 
on the continuum space. Consequently,
our mean field Hamiltonian is close to Hofstadter Hamiltonian and 
hence there are similarities between their solutions.

Hofstadter Hamiltonian shows remarkable structures. 
As is seen in Fig.1 the largest gap exists along a line 
$\Phi=\nu\Phi_0$ with a unit of flux $\Phi_0$. 
Ground state energy becomes minimum also with this flux. 
These facts may suggest that Hofstadter problem has some connection 
with the fractional Hall effect. 
We pursue a mean field theory of the condensed flux states in the 
quantum Hall system and point out that the Hofstadter problem is 
actually connected with the fractional Hall effect. 

$ $From the dynamical magnetic field, are defined new Landau levels. 
If integer number of these Landau levels are filled completely, 
the integer quantum Hall effect occurs. 
The ground state has a large energy gap and is stable against 
perturbations, just like ordinary integer quantum Hall effect. 
We study these states and will identify them as fractional quantum Hall 
states. 

We postulate, in the quantum Hall system of the filling factor $\nu$, 
the dynamical flux per plaquette and dynamical magnetic field of 
the following magnitudes,
\begin{eqnarray}
&\Phi_{\rm ind}=\nu\Phi_0,\ \Phi_0=\Phi_{\rm external\ flux},\\
&B_{\rm ind}=\nu B_0,\ B_0=B_{\rm external\ magnetic\ field},
\nonumber
\end{eqnarray}
where $\nu$ is the filling factor measured with the external magnetic 
field. 
We obtain a self-consistent solution with this flux. 
Then the integer quantum Hall effect due to induced magnetic field 
could occur just at filling factor $\nu$, because the density 
satisfies the integer Hall effect condition, 
\begin{eqnarray}
&{eB_{\rm ind}\over 2\pi}N={eB_0\over 2\pi}\nu,\\
&N=1.\nonumber
\end{eqnarray}
The ground state has a large energy gap, generally. 

At the half-filling $\nu=1/2$, half-flux $\Phi_0/2$ is induced. 
This situation has been studied in detail by Lieb\cite{r13} and 
others\cite{r14} in 
connection with Hubbard model or t-J model. 
Lieb gave a quite general proof that the energy is optimal with 
half-flux at the half-filling case. 
We study first the state of $\nu=1/2$ and the states of  
$\nu=p/(2p\pm1)$, next. 
At $\Phi=\Phi_0/2$, band structure is that of massless Dirac field and has 
doubling symmetry. 
When even number of Landau levels of the effective magnetic field, 
$B_{\rm ind}-B_0/2$, are filled, ground states have large energy gaps. 
This occurs if the condition of the density, 
\begin{eqnarray}
{e\over 2\pi}\vert\nu-{1\over2}\vert B_0\cdot 2p={eB_0\over2\pi}\nu,\\
\nu={p\over 2p\pm1}\ ;\ p,\ {\rm integer},\nonumber
\end{eqnarray}
is satisfied. A factor 2 in the left-hand side is due to doubling of 
states and will be discussed later.
We study these states in detail based on von Neumann lattice representation. 

Action, Eq.(2.11), and density operator, Eq.(2.12), show that 
there is an effective  magnetic field in the momentum space. 
Area of the momentum space is given by a finite value, $(2\pi/a)^2$, 
and the total flux is hence finite. 
The total flux in the momentum space is in fact unit flux. 
In the thermodynamic limit, in which the density in space is finite, 
the density in momentum space is infinite. 
Consequently, it is possible to make this phase factor disappear by 
a singular gauge transformation in the momentum space with 
infinitesimally small coupling. 

We make a singular gauge transformation of the field in the momentum 
space, 
\begin{eqnarray}
&c_l({\bf p})=e^{i\tilde e\lambda({\bf p})}b_l({\bf p}),
\nonumber\\
&\lambda({\bf p})={1\over 2\pi}\int\theta({\bf p}-{\bf p}')\rho({\bf p}')
d{\bf p}',\\
&\tan\theta={(p-p')_y\over(p-p')_x},\nonumber
\end{eqnarray}
where $\rho({\bf p})$ is the density operator in the momentum space and 
$\tilde e$ is determined from Eqs.(B.3) and (B.4) in Appendix B. 
With the transformed field, the commutation relation and the 
charge density are expressed as, 
\begin{eqnarray}
&\{c_{l_1}({\bf p}_1),c^\dagger_{l_2}({\bf p}_2)\}=
\sum\delta_{l_1,l_2}\delta({\bf p}_1-{\bf p}_2+{2\pi\over a}{\bf N}),\\
&\rho({\bf k})=\sum c^\dagger_{l_1}({\bf p}_1)c_{l_2}
({\bf p}_2)\delta({\bf p}_1-{\bf p}_2-{\bf k}+{2\pi\over a}{\bf N})
(l_1\vert e^{i{\bf k}\cdot\hbox{\bxi}}\vert l_2).\nonumber
\end{eqnarray}
By a Chern-Simons gauge theory in the momentum space, the gauge 
transformation, Eq.(3.4), is realized as is expressed in Appendix B. 
Here, the coupling constant $\tilde e$ is infinitesimally 
small, hence fluctuations of the Chern-Simons gauge field have small 
effect and we ignore the fluctuations. 

The action in the coordinate representation is given by, 
\begin{eqnarray}
&S=\sum c^\dagger_l({\bf R})(i\hbar{\partial\over\partial t}-E_l)
c_l({\bf R})-{1\over 2}\sum v_{l_1,l_2;l_3,l_4}({\bf R}_2-{\bf R}_1)
c^\dagger_{l_1}({\bf R}_1)c^\dagger_{l_2}({\bf R}_1)
c_{l_3}({\bf R}_2)c_{l_4}({\bf R}_2),\nonumber\\
&v_{l_1,l_2;l_3,l_4}({\bf R}_2-{\bf R}_1)=\int_{{\bf k}\neq0}
d{\bf k} V({\bf k})
(l_1\vert e^{i{\bf k}\cdot\hbox{\bxi}}\vert l_2)
(l_3\vert e^{-i{\bf k}\cdot\hbox{\bxi}}\vert l_4)
e^{i{\bf k}\cdot({\bf R}_2-{\bf R}_1)}.
\end{eqnarray}
Hence the Hamiltonian in the lowest Landau level space is given by,
\begin{eqnarray}
&H=-{1\over 2}\sum v({\bf R}_2-{\bf R}_1) c_0^\dagger({\bf R}_1)
c_0({\bf R}_2) c_0^\dagger({\bf R}_2) c_0({\bf R}_1),
\\
&v({\bf R})=
{\pi\over a}e^{-{\pi\over2}{\bf R}^2}I_0({\pi\over2}{\bf R}^2),
\nonumber
\end{eqnarray}
where $I_0$ is zero-th order modified Bessel function. 
We study a mean field solution of this Hamiltonian. 
We have an expectation value and a mean field Hamiltonian, 
\begin{eqnarray}
&\langle c_0^\dagger({\bf R}_1) c_0({\bf R}_2)\rangle=
U_0({\bf R}_1-{\bf R}_2)e^{i\int_{{\bf R}_2}^{{\bf R}_1}{\bf A}_{\rm ind}
\cdot d{\bf x}},\\
&H_{\rm mean\ field}=-\sum v({\bf R}_2-{\bf R}_1)U_0({\bf R}_1-{\bf R}_2)
e^{i\int_{{\bf R}_2}^{{\bf R}_1}{\bf A}_{\rm ind}\cdot d{\bf x}}
c_0^\dagger({\bf R}_2)c_0({\bf R}_1),
\nonumber
\end{eqnarray}
and solve the equations self-consistently. 
The mean field Hamiltonian coincides to that of Hofstadter if the 
potential is of short range of nearest neighbor type. 
The spectrum obtained by Hofstadter shows characteristic structures, 
and has a deep connection with the structure of the fractional quantum 
Hall effect. 

\ 

(i) Half-filled case, $\nu=1/2$.

At half-filling $\nu=1/2$, the system has a half flux 
$\Phi=\Phi_0/2$. 
The system, then, is described equivalently with the two-component 
Dirac field by combining the field at even sites with that at 
odd sites. 
In the gauge, $A_x=0,\ A_y=Bx$, the mean field Hamiltonian reads, 
\begin{eqnarray}
&H_{\rm M.F.}=
\sum\Psi^\dagger(X')
\left( \begin{array}{cc}
a_{ee}(X'-X) & a_{eo}(X'-X)\\
a_{oe}(X'-X) & a_{oo}(X'-X)
\end{array}\right)\Psi(X),
\nonumber\\
&\Psi(X)=
\left(\begin{array}{c}
c(2m,n)\\
c(2m+1,n)
\end{array}\right),
\end{eqnarray}
\begin{eqnarray}
&a_{ee}(m'-m,n'-n)=\langle c^\dagger(2m',n')c(2m,n)\rangle
v(2m'-2m,n'-n),\nonumber\\
&a_{oo}(m'-m,n'-n)=\langle c^\dagger(2m+1',n')c(2m+1,n)\rangle
v(2m'-2m,n'-n),\nonumber\\
&a_{eo}(m'-m,n'-n)=\langle c^\dagger(2m',n')c(2m+1,n)\rangle
v(2m'-2m-1,n'-n),\nonumber\\
&a_{oe}(m'-m,n'-n)=\langle c^\dagger(2m'+1,n')c(2m,n)\rangle
v(2m'-2m+1,n'-n),\nonumber\\
&a_{eo}=a_{oe},\\
&a_{ee}+a_{oo}=0.\nonumber
\end{eqnarray}
We obtain the self-consistent solutions numerically. 
Fig.2 shows the spectrum. 
As is expected, spectrum has two minima and two zeros corresponding 
to doubling. 
We have the momentum space expression of the mean field Hamiltonian, 
\begin{eqnarray}
&H_{\rm M.F.}&=\sum c^\dagger_\xi({\bf p})F_{\xi\eta}({\bf p})c_\eta
({\bf p})\\
&&=\sum\epsilon_\alpha({\bf p})c^\dagger_\xi({\bf p})U^\dagger
_{\xi\xi'}({\bf p})U_{\xi'\eta}({\bf p})c_\eta({\bf p}),\nonumber\\
&&\epsilon_\pm({\bf p})=\pm\sqrt{a_{ee}^2({\bf p})+a_{eo}^2({\bf p})},
\nonumber\\
&&U_{\xi\eta}({\bf p})=
\left(\begin{array}{cc}
a_{eo}/N_+ & (\epsilon_+ -a_{ee})/N_+\\
a_{eo}/N_- & (\epsilon_- -a_{ee})/N_-
\end{array}\right),\nonumber\\
&&N^2_{\pm}=2(a^2_{ee}+a^2_{eo}-a_{ee}\epsilon_{\pm}),
\nonumber
\end{eqnarray}
$$
a_{ee}(p_x,p_y+\pi/a)=-a_{ee}(p_x,p_y),\ 
a_{ee}(p_y,p_x)=a_{eo}(p_x,p_y).\nonumber
$$

The  matrix $F_{\xi\eta}({\bf p})$ is approximated well with a nearest 
neighbor form,
\begin{equation}
F({\bf p})=2K\left(\begin{array}{cc}
\cos p_y&\cos p_x\\
\cos p_x&-\cos p_y
\end{array}\right),
\end{equation}
$$
K=0.107{e^2\over\kappa l_B},\ l_B=\sqrt{\hbar\over eB}.
$$
Around minima, the energy eigenvalue of Eq.(3.11) is approximated as,
\begin{equation}
E(p)=E_0+{({\bf p}-{\bf p}_0)^2\over 2m^*},\ 
{\bf p}_0=(0,0),(0,\pi/a),
\end{equation}
and they are approximated around zeros as,
\begin{eqnarray}
&E(p)=\gamma{\hbox{\bal}}\cdot{\bf p},\\
&\alpha_x=\left(\begin{array}{cc}
0&1\\
1&0
\end{array}\right),\ 
\alpha_y=\left(\begin{array}{cc}
1&0\\
0&-1
\end{array}\right),
\nonumber
\end{eqnarray}
in $2\times2$ expression. 
The $m^*$ in Eq.(3.13) is the effective mass and $\gamma$ is the effective 
velocity. 
They are computed numerically. 
Its value is, 
\begin{eqnarray}
&m^*=0.225\sqrt{B\over B_0}m_e,\ B_0=20{\rm Tesla},\\
&\gamma=0.914{e^2\over\kappa}.\nonumber
\end{eqnarray}
The effective mass is proportional to the square root of 
$B$ in our method. 

$\nu=1/2$ mean field Hamiltonian is invariant under a kind of 
Parity, $P$, and 
anti-commutes with a  chiral transformation, $\alpha_5$, which are 
defined by,
\begin{eqnarray}
P:\ &\Psi(p_x,p_y)&\rightarrow \alpha_x\Psi(p_x,p_y+\pi/a),\\
\alpha_5:\ &\Psi(p_x,p_y)&\rightarrow\alpha_5\Psi(p_x,p_y),\nonumber\\
&&\alpha_5=\alpha_x\alpha_y.\nonumber
\end{eqnarray}
If the parity is not broken spontaneously, there is a degeneracy 
due to parity doublet. 
Doubling of the states\cite{r15} appears also at $\nu\neq1/2$ and plays 
important role, 
when we discussed the states away from $\nu=1/2$ in the next part. 
When an additional vector potential with the same gauge, 
$A_x=0,\ A_y=Bx$, is added, the Hamiltonian satisfies the  properties 
under the above transformations and 
the doubling due to parity doublet also appears. Thus,
the factor 2 is necessary in Eq.(3.3) and leads the principal 
series at $\nu=p/(2p\pm1)$ to have maximum energy gap. 

\ 

(ii) $\nu={p\over 2p\pm1}$

If the filling factor, $\nu$, is slightly away from 1/2, total system 
can be regarded as a system with a small magnetic field 
$(\nu-1/2)B_0$. 
A band structure may be slightly modified. 
It is worthwhile to start from the band of $\nu=1/2$ as a first 
approximation and to make iteration in order to obtain self-consistent 
solutions at arbitrary $\nu=p/(2p\pm1)$. 

We solve the following mean field Hamiltonian under the self-consistency 
condition at $\nu=1/2+\delta$, 
\begin{eqnarray}
&H_{\rm M}=\sum U^{({1\over 2}+\delta)}_0({\bf R}_1-{\bf R}_2)
e^{i\int({{\bf A}}^{({1\over 2})}+
\delta{\bf A})d{\bf x}}v({\bf R}_1-{\bf R}_2)
c^\dagger({\bf R}_1)c({\bf R}_2),\\ 
&\langle c^\dagger({\bf R}_1)c({\bf R}_2)\rangle_{1/2+\delta}=
U^{({1\over 2}+\delta)}_0 e^{i\int({{\bf A}}^{({1\over 2})}
+\delta{\bf A})}.\nonumber
\end{eqnarray}
Here we solve, instead, an Hamiltonian which has the 
phase of Eq.(3.17) but has the magnitude of the $\nu=1/2$ state. 
Namely we study a series of states defined at $\nu=1/2$,
\begin{eqnarray}
H_{\rm M}^{(2)}&=\sum U^{({1\over2})}_0({\bf R}_1-{\bf R}_2)
e^{i\int{\bf A}^{({1\over2})}d{\bf x}}v({\bf R}_1-{\bf R}_2)
e^{i\int\delta{\bf A}d{\bf x}}c^\dagger({\bf R}_1)c({\bf R}_2)
\nonumber\\
&=\sum F^{({1\over2})}({\bf R}_1,{\bf R}_2;\delta{\bf A})
c^\dagger({\bf R}_1)c({\bf R}_2).
\end{eqnarray}
Integer quantum Hall state has an energy gap 
of the Landau levels due to $\delta{\bf A}$. 
This occurs when the integer number of Landau levels are filled 
completely. 
Landau level structure is determined by the phase factor and 
Eq.(3.17) and Eq.(3.18) have the same Landau level structure. 
Magnitudes of the physical quantities of Eq.(3.17) may be modified, 
nevertheless.

$F^{({1\over2})}({\bf R}_1,{\bf R}_2;0)$ was obtained in the previous part, 
and is approximated with either the effective mass formula (3.13) or 
with the nearest neighbor form (3.12). 
$ $From Eqs.(3.13) and (3.18), 
the Hamiltonian $H^{(2)}_{\rm M}$ is written in the former method as 
\begin{eqnarray}
H_{\rm M}^{(2)}&=\sum c^\dagger({\bf R}_1)U^\dagger({\bf R}_1,
{\bf R}'_1,\delta{\bf A})
\{E_0+{({\bf p}+e\delta{\bf A})^2\over2m^*}\}
U({\bf R}'_2,{\bf R}_2,\delta{\bf A})c({\bf R}_2)\nonumber\\
&=\sum\tilde c^\dagger(R_1)
\{E_0+{({\bf p}+e\delta{\bf A})^2\over2m^*}\}\tilde c({\bf R}_1),\\
&\tilde c({\bf R}_1)=\sum U({\bf R}_1,{\bf R}'_1)c({\bf R}'_1),\\
&U_{\xi\eta}({\bf R}_1,{\bf R}'_1)=\int d{\bf p} e^{i{\bf p}\cdot({\bf R}_1-
{\bf R}'_1)}U_{\xi\eta}({\bf p}).
\end{eqnarray}
Integer Hall states of Eq.(3.19) satisfy, 
\begin{equation}
\langle{\tilde c}^\dagger_\xi({\bf R}_1)
{\tilde c}_\eta({\bf R}_2)\rangle=
\tilde u_{\xi\eta}({\bf R}_1-{\bf R}_2)
e^{i\int^{{\bf R}_1}_{{\bf R}_2}\delta{\bf A}d{\bf x}},
\end{equation}
and leads the expectation value, 
\begin{eqnarray}
\langle c^\dagger({\bf R}_1)c({\bf R}_2)\rangle&=&U^\dagger(
{\bf R}_1,{\bf R}'_1)
\langle{\tilde c}^\dagger({\bf R}'_1){\tilde c}({\bf R}'_2)\rangle
U({\bf R}'_2,{\bf R}_2)\\
&=&U^\dagger({\bf R}_1,{\bf R}'_1)\tilde u_{\xi\eta}({\bf R}'_1-{\bf R}'_2)
e^{i\int^{{\bf R}'_1}_{{\bf R}'_2}\delta{\bf A}d{\bf x}}
U({\bf R}'_2,{\bf R}_2).\nonumber
\end{eqnarray}

Here, we solve the equations obtained from Eq.(3.19) with continuum 
approximation, first. 
We have, 
\begin{eqnarray}
&[E_0+{({\bf p}-{\bf p}_0^{(i)}+\delta{\bf A}_{\rm ind})^2\over 2m^*}]
u^{(i)}_p=E_p u_p^{(i)},\\
&i=1,2,\nonumber\\
&E_p=E_0+{e\delta B_{\rm eff}\over 2m^*}(2p+1)\\
&\delta B_{\rm eff}=\vert \nu-{1\over2}\vert B_0.\nonumber
\end{eqnarray}
$ $From Eq.(3.25), 
$p$ Landau levels are completely filled and the integer 
quantum Hall effect occurs at $\nu=p/(2p\pm1)$. 
The energy gap is given by Landau level spacing, 
\begin{equation}
\Delta E_{\rm gap}={e\delta B_{\rm eff}\over m^*}={eB_0\over m^*}
\vert\nu-{1\over2}\vert. 
\end{equation}

Hamiltonian with the nearest neighbor form is a $2\times2$ matrix of the 
tight-binding type. 
Based on this Hamiltonian, we solve Landau level equation of 
the tight-binding  form  with the vector potential 
$\delta{\bf A}$, 
\begin{eqnarray}
&K\{c_1({\bf n}+{\bf\Delta}_x)e^{i\int_{\bf n}^{{\bf n}+{\bf\Delta}_x}
\delta{\bf A}\cdot d{\bf x}}+
c_2({\bf n}+{\bf \Delta}_y)e^{i\int_{\bf n}^{{\bf n}+{\bf\Delta}_y}
\delta{\bf A}\cdot d{\bf x}}+{\rm h.c.}\}=Ec_1({\bf n}),\nonumber\\
&K\{c_1({\bf n}+{\bf\Delta}_x)e^{i\int_{{\bf n}}^{{\bf n}+{\bf\Delta}_x}
\delta{\bf A}\cdot d{\bf x}}-
c_2({\bf n}+{\bf\Delta}_y)e^{i\int_{\bf n}^{{\bf n}+{\bf\Delta}_y}
\delta{\bf A}\cdot d{\bf x}}+{\rm h.c.}\}=Ec_2({\bf n}).\nonumber\\
&
\end{eqnarray}
The equation are solved numerically and the energy gaps and the 
widths of excited bands are given in Fig.3. 
Some bands are narrow and some bands are wide. 
Near $\nu=1/2$, the effective magnetic field approaches to zero and 
Landau level wave functions have large spatial extensions. 
Lattice structure becomes negligible and spectrum shows simple 
Landau levels of the continuum equation in these regions. 
Near $\nu=1/3$, lattice structure is not negligible and bands have 
finite widths. 
There are non-negligible corrections from those of continuum 
calculations, Eq.(3.24).

Due to energy gap of the integer Hall effect caused by the induced 
dynamical magnetic field, the states at $\nu=p/(2p\pm1)$ are stable 
and fluctuations are weak. 
Invariance under $P$, moreover, ensures these states to 
have uniform density. 
In systems with impurities, localized states with isolated discrete 
energies are generated by impurities and have energies in the gap regions. 
They contribute to the density but do not contribute to the conductance. 
If the Fermi energy is in one of these gap regions, 
the Hall conductance $\sigma_{xy}$ is given by a topological formula, 
Eq.(2.19), and stays constant, at 
${e^2\over h}\cdot{p\over 2p\pm1}$. The fractional Hall effect is realized.
 
At a value of $\nu$ smaller than 1/3, 
Hofstadter butterfly shows other kind of 
structures hence suggests other structures of the fractional Hall effect do 
exist. 
For instance at $\nu=$1/5, 1/7, 1/9,$\dots$the energy gap is large. 
In low density, it may be complicated, in fact. 
There could exist completely 
different kind of phase, such as Wigner crystal phase\cite{r16}. 
Competition between two phases may be important. 
They will not be presented in 
the present paper but it will be presented in a later work. 

\ 

(iii) Fluctuations of FQHE at $\nu={p\over 2p\pm1}$.

Ground states have energy gap, hence the fluctuations are small, just as 
in the integer Hall effect. 
Density and phase fluctuations are described by the massive 
Chern-Simons gauge theory\cite{r17}with a mass of order energy gap. 

\ 

(iv) Fluctuations at half-filling, $\nu={1\over2}$.

Ground state has no energy gap at $\nu=1/2$. 
Fluctuations are described by the action, 
\begin{equation}
S=\int d^3x\Psi^\dagger(i\hbar{\partial\over\partial t}+a_0)\Psi+
\gamma\Psi^\dagger\hbox{\bal}\cdot({\bf p}+{\bf a})\Psi
\end{equation}
The fermion field integration leads severe infra-red divergence. 
If the energy dispersion is changed to 
\begin{equation}
E_0=\tilde\gamma\vert p\vert^{1+\delta},
\end{equation}
by interaction, physics at $\nu=1/2$ is completely different from that 
of the mean field. 
We will not discuss physical properties of $\nu=1/2$ state in this paper.

\section{Comparison with experiments}\scs

In the previous section we presented our mean field theory based on 
flux condensation, where lattice structure generated by the external magnetic  
field and condensed flux due to interaction 
are important ingredient. 
Consequently, our mean field Hamiltonian becomes very similar to 
that of Hofstadter which is known to show actually large energy gap 
zone along $\Phi=\nu\Phi_0$ line. 
The line $\Phi=\nu\Phi_0$ is special in Hofstadter problem 
and hence in our mean field Hamiltonian, too. 
This explains why the experiments of the fractional quantum Hall effects 
shows characteristic behavior at $\nu=p/(2p\pm1)$. 
The ground states at $\nu=p/(2p\pm1)$ have lowest energy 
and largest energy gap, hence these states are 
stable. 
In this section we compare the energy gaps of the principal series 
with the experiments in the lowest order. 
and the ground state energy of $\nu=1/3$ state with the Laughlin 
variatioal wave function\cite{r18}.

The effective mass $m^*$ of Eq.(3.14) was obtained from the curvature 
of the energy dispersion and should show a characteristic mass 
scale of the fractional Hall effect. 
Eq.(3.24) gives Landau level energy in the lowest approximation and 
the gap energy is given in Eq.(3.25). 
The gap energy from the nearest neighbor approximation, Eq.(3.12) is 
given by solving Eq.(3.27). 
They are compared with the experimental values\cite{r19} 
in Fig.4. 
The agreement is not perfect but should be regarded good as the 
lowest mean field approximation. 
Near $\nu=1/3$, the bands have finite widths and near $\nu=1/2$, the 
widths are infinitesimal. 
The dependence of the width upon the filling factor, $\nu$, 
and the whole structure of the bands expressed in Fig.1 are 
characteristic features of the present mean field and 
should be tested experimentally. 

Finally we compute the ground state energy of the $\nu=1/2$ state and 
the $\nu=1/3$ state. 
Using the fact that our mean field Hamiltonian of $\nu=1/3$ is very 
close to the short-range tight-binding Hamiltonian, 
we compute the ground state energy of $\nu=1/3$ with the tight-binding 
model wave function. 
The wave function is obtained numerically and is substituted 
to the total energy per particle, Eq.(3.7), as 
\begin{eqnarray}
E_{1/3}&=&{3\over N}\langle\Psi\vert H\vert\Psi\rangle \\
&=&-{3\over2}\sum_{\bf X}v({\bf X})\vert U_0({\bf X})\vert^2
.\nonumber
\end{eqnarray}
where $N$ is the number of sites. 
The result is,
\begin{equation}
E_{1/3}=-0.340{e^2\over\kappa l_B}.
\end{equation}
where $l_B=\sqrt{\hbar/eB}$. 
This value should be compared with that of Laughlin wave function,
\begin{equation}
E_{1/3}=-0.416{e^2\over\kappa l_B}.
\end{equation}
The value of Eq.(4.2) is higher than Eq.(4.3), but the difference is 
not large. 
This may suggest that mean field flux state is infact close to 
Laughlin wave function and to the exact solution. 
For the state at $\nu=1/2$, we use our self-consistent solution 
for computing the ground state energy per particle, 
\begin{equation}
E_{1/2}=-0.347{e^2\over\kappa l_B}.
\end{equation}
As is mentioned in Section.3, the fluctuations are extremely large and 
the mean field value get large corrections from the higher order 
effects. 
So the results about the $\nu=1/2$ state should not be 
taken seriously.

\section{Summary}\scs

We formulated the quantum Hall effects, integer Hall effect and 
fractional Hall effect with von Neumann lattice representation of 
two-dimensional electrons in a strong magnetic field. 
von Neumann lattice is a subset of coherent state. 
The overlapp of the states is expressed with theta function. 
They give a systematic way of expressing quantum Hall dynamics. 

Topological invariant expression of the Hall conductance was 
obtained in which compactness of the momentum space is 
ensured by the lattice of the coordinate space. 
Because the lattice has an origin in the external magnetic field, 
topological character of the Hall conductance is ensured by 
the external magnetic field. 
The conductance is quantized exactly as $(e^2/h)\cdot N$ at the 
quantum Hall regime. 
The integer $N$ increases with chemical potential. 

A new mean field theory of the fractional Hall effect that has 
dynamical flux condensation is studied. 
If the filling $\nu$ is less than one, many particle states 
of Landau levels have no energy gap unless interaction switchs on. 
In the tight-binding model, the situation is very different. 
The spectrum that was found by Hofstadter first is changed drastically 
when the flux is changed and it has a 
large energy gap in some regions. 
In our mean field theory, lattice structure is introduced from von 
Neumann lattice and flux is introduced dynamically. 
The mean field Hamiltonian becomes a kind of tight-binding model and 
rich structure of the tight-binding model is seen, in fact, as 
characteristic features of the fractional Hall effect. 
Our mean field flux states have liquid property of uniform density with 
energy gap. 
They are defined as special integer quantum Hall states 
in the lowest Landau level space hence the band structures shown in 
Figs.2 and 3 are very different from those of normal integer Hall states. 
We gave a dynamical reason why the principal series at $\nu=p/(2p\pm1)$ are 
observed dominantly. 
These states satisfy the self-consistency condition of having 
the lowest energy and the largest energy gap. 
The physical quantities of our mean field theory are close to the 
experimental values in the lowest order at $\nu=p/(2p\pm1)$. 
At the exact half-filling $\nu=1/2$, fluctuations are very large and
corrections from mean field value may be large, too. 

\section*{Acknowledgements}
We are indebted to Professors P. Wiegmann and H. Suzuki for their 
useful comments in the early stage of the present work. 
One of the authors(K.I.) thanks Professors A. Luther and 
H. Nielsen for useful discussions. 
The present work was partially supported by the special Grant-in-Aid 
for Promotion of Education and Science in Hokkaido University 
Provided by the Ministry of Education, Science, Sports, and Culture, 
a Grant-in-Aid for Scientific Research(07640522), and 
Grant-in-Aid for International Scientific Research(Joint Research
07044048) the Ministry of Education, Science, Sports, and Culture, 
Japan.

\newpage
\renewcommand{\theequation}{A.\arabic{equation}}
\noindent {\Large \bf APPENDIX A}
\setcounter{equation}{0}
\bigskip

In the present representation, a short-range impurity term can be 
expressed as, 
\begin{eqnarray}
&H_{\rm impurity}=a^4\int^{\pi/a}_{-\pi/a}{d^2p_1\over (2\pi)^2}
{d^2p_2\over (2\pi)^2}b^\dagger(p_1)\sum_{{\bf N}}({2\pi\over a})^2 
e^{ia\int^{p_2^N}_{p_1}{\bf A}\cdot d{\bf p}}\tilde V(p_1-p_2^N)e^{
-{a^2(p_1-p_2^N)^2\over8\pi}}b(p_2^N)\nonumber\\
&=a^2\int^{\pi/a}_{-\pi/a}{d^2p_1\over (2\pi)^2}b^\dagger(p_1)\int^\infty
_{-\infty}d^2 k\tilde V(-k)e^{-{a^2k^2\over8\pi}}e^{iak^i D_i}b(p_1),
\nonumber\\
&D_i={1\over i}{\partial\over a\partial p_i}+A_i,\ 
{\bf A}=({ap_y\over2\pi},0),
\\
&{\bf p}_2^N={\bf p}_2+2\pi{\bf N},\ 
V(x)=\int^\infty_{-\infty}d^2k\tilde V(k)e^{i{\bf k}\cdot{\bf x}}.
\nonumber
\end{eqnarray}
With the creation and annihilation operator defined by,
\begin{eqnarray}
&A=D_x-iD_y,\\
&A^\dagger=D_x+iD_y,\nonumber\\
&[A,A^\dagger]={1\over\pi},\nonumber
\end{eqnarray}
the above Hamiltonian can be written as,
\begin{equation}
H_{\rm impurity}=a^2\int^{\pi/a}_{-\pi/a}{d^2p\over (2\pi)^2}
b^\dagger(p)\int^\infty_{-\infty}d^2 k\tilde V(-k)e^{{a\over2}(ik_x-k_y)A}
e^{{a\over2}(ik_x+k_y)A^\dagger}b(p).
\end{equation}
We represent $H_{\rm impurity}$ in Landau level representation of 
the momentum space defined by,
\begin{eqnarray}
b(p)&=&\sum_l b_l\tilde\psi_l(p),\ 
\vert l\rangle=\pi^{l\over2}{1\over \sqrt{l!}}
(A^\dagger)^l\vert 0\rangle,\\
A\vert 0\rangle&=&0,\nonumber\\
\tilde\psi_l(p)&=&\langle p\vert l\rangle
=C\sum_N e^{-i\pi N}e^{ik_N a p_x}H_l({a(p_y+{2\pi\over a} 
k_N)\over\sqrt{\pi}})e^{-{a^2\over4\pi}(p_y+{2\pi\over a} k_N)^2},
\nonumber
\end{eqnarray}
where $k_N=N+{1\over2}$ and $C$ is the normalization constant. 
The normalized lowest Landau level wave function is,
\begin{equation}
\tilde\Psi_0=2^{1/4}e^{-{a^2 p_y^2\over4\pi}}
\Theta_1({a(p_x+ip_y)\over2\pi},i).
\end{equation}
$H_{\rm impurity}$ is reduced to,
\begin{eqnarray}
&H_{\rm impurity}=\sum_{l_1,l_2}b^\dagger_{l_1}\tilde V_{l_1,l_2}
b_{l_2},\\
&\tilde V_{l+n,l}=4\pi\sqrt{l!\over (l+n)!}\int d^2 k
\tilde V(-2\sqrt{\pi} k)a^n (ik_x+k_y)^n L_l^{(n)}
(k^2 a^2)e^{-a^2 k^2},\nonumber\\
&\tilde V_{l,l+n}=4\pi\sqrt{l!\over (l+n)!}\int d^2 k
\tilde V(-2\sqrt{\pi} k)a^n (ik_x-k_y)^n L_l^{(n)}
(k^2 a^2)e^{-a^2 k^2}.\nonumber
\end{eqnarray}
For a short range potential, we have
\begin{eqnarray}
&V(x)=g\delta(x),\ 
\tilde V(-k)={g\over (2\pi)^2},\nonumber\\
&\tilde V_{l+n,l}=\tilde V_{l,l+n}=0,\ {\rm for}\ 
n\neq0,\\
&\tilde V_{l,l}=0,\ {\rm for}\ l\neq0,\nonumber\\
&\tilde V_{0,0}={g\over (2\pi)^2}4\pi\int d^2 k 
L_0^{(0)}(a^2k^2)e^{-a^2k^2}={g\over a^2},\nonumber\\
&H_{\rm impurity}=\tilde V_{0,0}b^\dagger_0 b_0.
\end{eqnarray}
Hence the one Landau level has the energy shift $\tilde V_{00}$, and 
all the other Landau levels have no effect from the impurity. 
The state $\vert l=0\rangle$ has an isolated energy and its wave function 
is square-integrable. 
This state corresponds to localized state. 

\newpage
\renewcommand{\theequation}{B.\arabic{equation}}
\noindent {\Large \bf APPENDIX B}
\setcounter{equation}{0}
\bigskip

In Chern-Simons gauge theory in momentum space, the action is
\begin{equation}
\int dtd{\bf p}
{1\over2}\epsilon_{\mu\nu\rho}a^\mu{\partial\over\partial p_\nu} 
a^\rho+a_0 j^0({\bf p})+
F(c^\dagger(p_1)e^{i\tilde e\int^{p_1}_{p_2}a_i dp_i}c
(p_2)),
\end{equation}
where in the last term operators $b^\dagger(p_1)b(p_2)$ are replaced with 
$c^\dagger(p_1)e^{i\tilde e\int^{p_1}_{p_2}a_i dp_i}c
(p_2)$, and the theory is defined on the torus\cite{r20}. 
The vector potential $a_i$ satisfy,
\begin{equation}
\epsilon_{0ij}{\partial\over\partial p_i} a^j+j^0({\bf p})=0.
\end{equation}
Its solution is substituted into the first term of (B.1). 
By choosing the coupling strength $\tilde e$ from the condition, 
\begin{eqnarray}
\tilde e j^0({\bf p})+{a^2\over 2\pi}=0,\\
j^0({\bf p})=({a\over 2\pi})^2 N_{\rm total}.
\end{eqnarray}
the phase factor which expresses the magnetic field in the momentum space 
in Eqs.(2.11), (2.12), (2.13), and (2.14) are cancelled. 
In thermodynamic limit, $j^0({\bf p})$ diverges, 
hence $\tilde e$ in infinitesimally small. 
The fluctuation of Chern-Simons gauge field can be ignored. 
The first two terms in Eq.(B.1) cancell if Eq.(B.2) is satisfied and 
do not contribute to the energy of the system.

\newpage
\centerline{\Large Figure Caption}
\noindent
Fig.1 : 
Hofstadter butterfly. 
The spectrum of the principal series are shown by thick lines. 

\noindent
Fig.2 : 
Spectrum of $\nu=1/2$ state, $\epsilon_-({\bf p})$. 
The unit is $(2/\pi a)(e^2/\kappa)[{\rm J}]$. 
 
\noindent
Fig.3 : 
Energy gaps(solid circles) and Band widths of excited states(open 
circles) of the principal series. 
The unit is $K$[J]. 

\noindent
Fig.4 : 
The energy gaps are calculated in the nearest neighbor approximation
(dots) and in the effective mass approximation(solid line). 
The experimental values are shown by the dashed line.

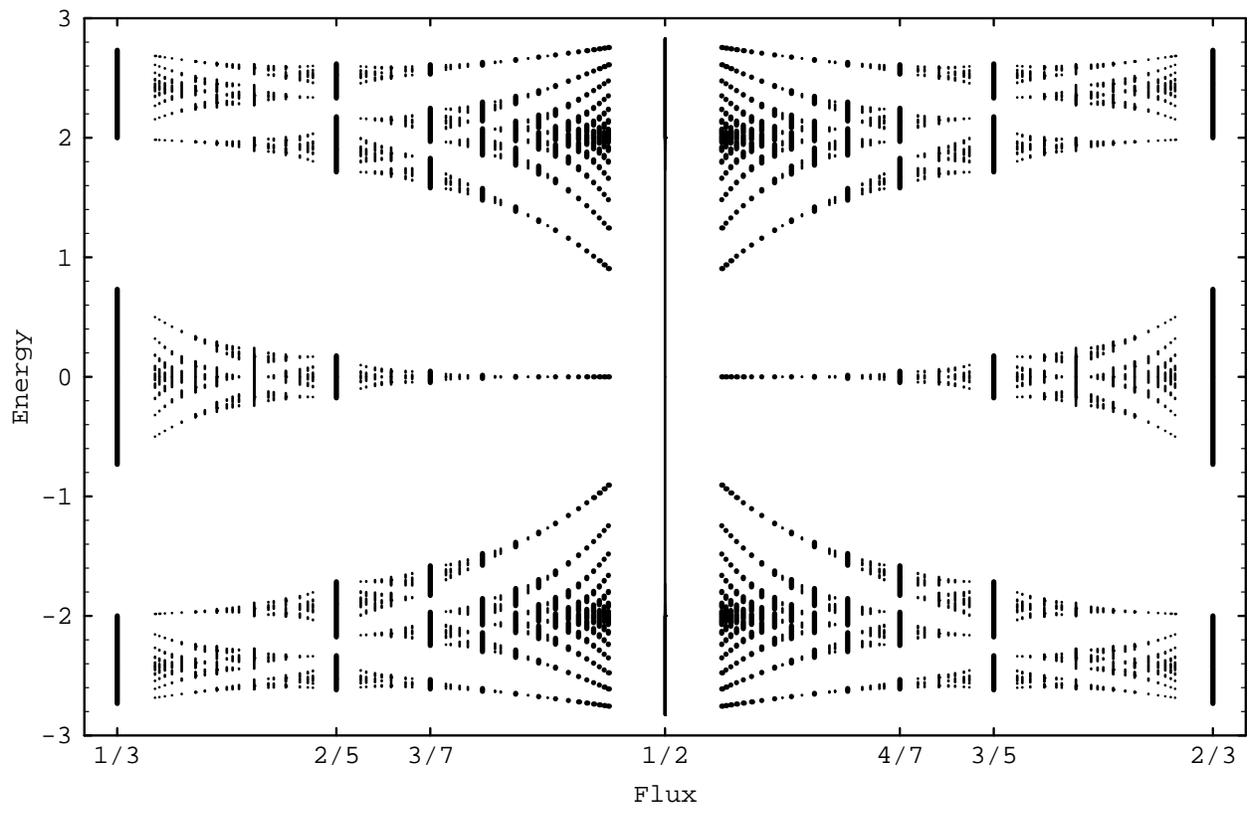
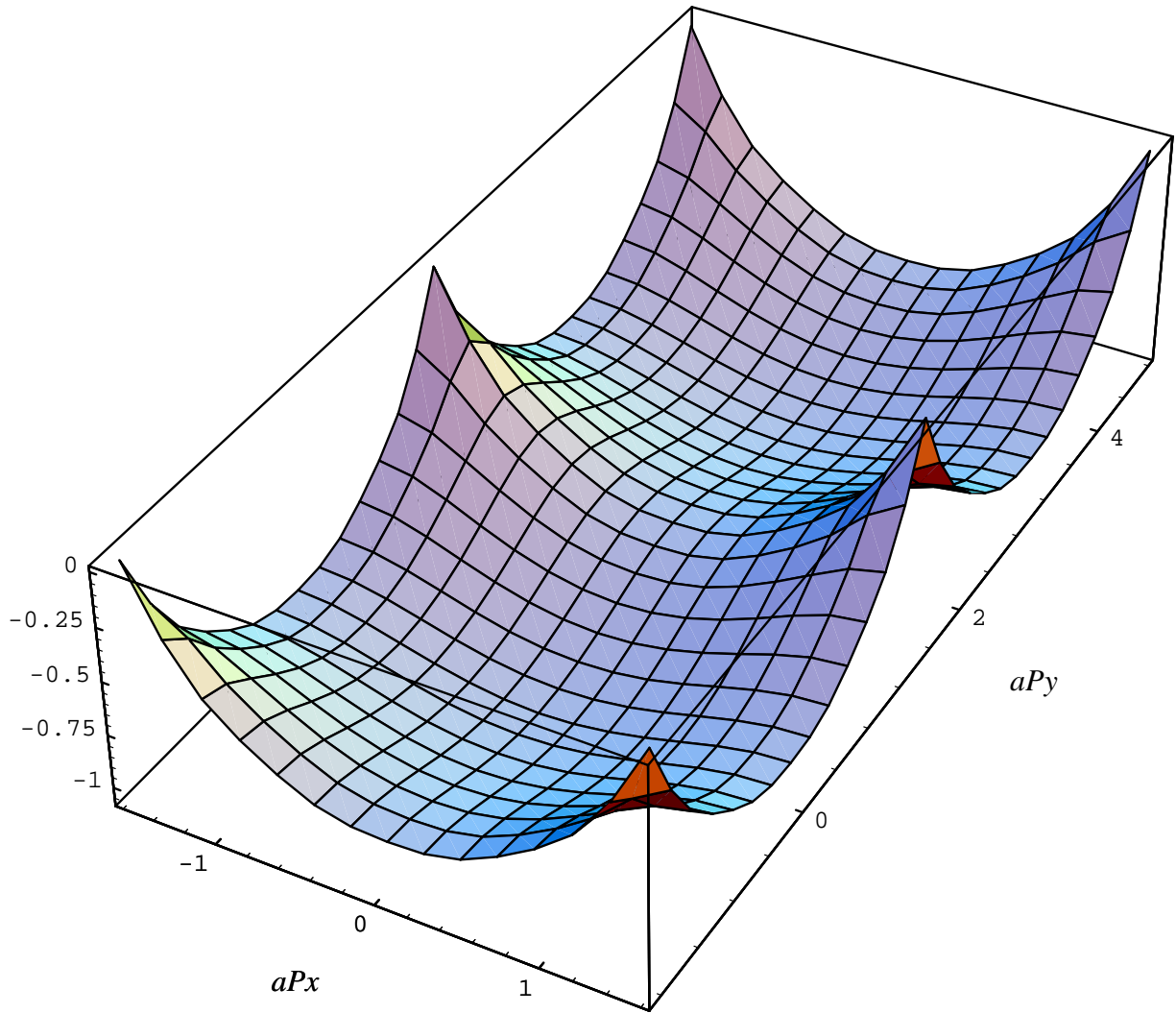
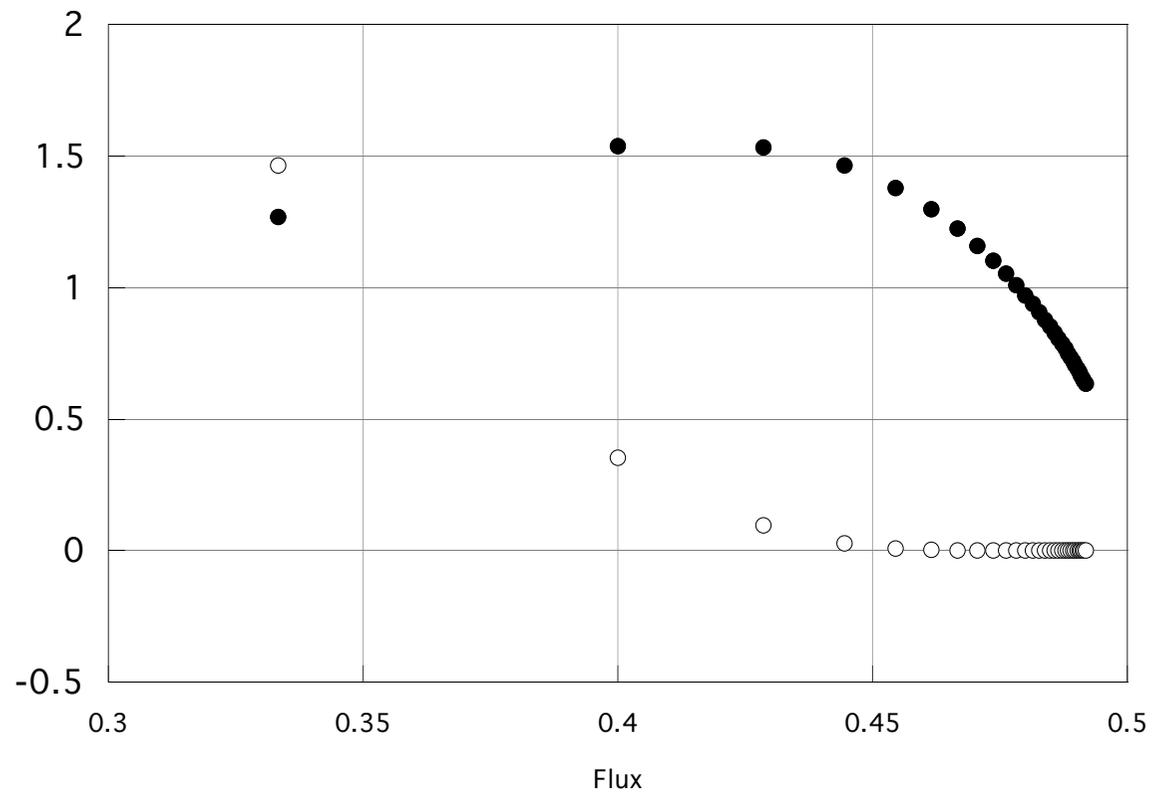
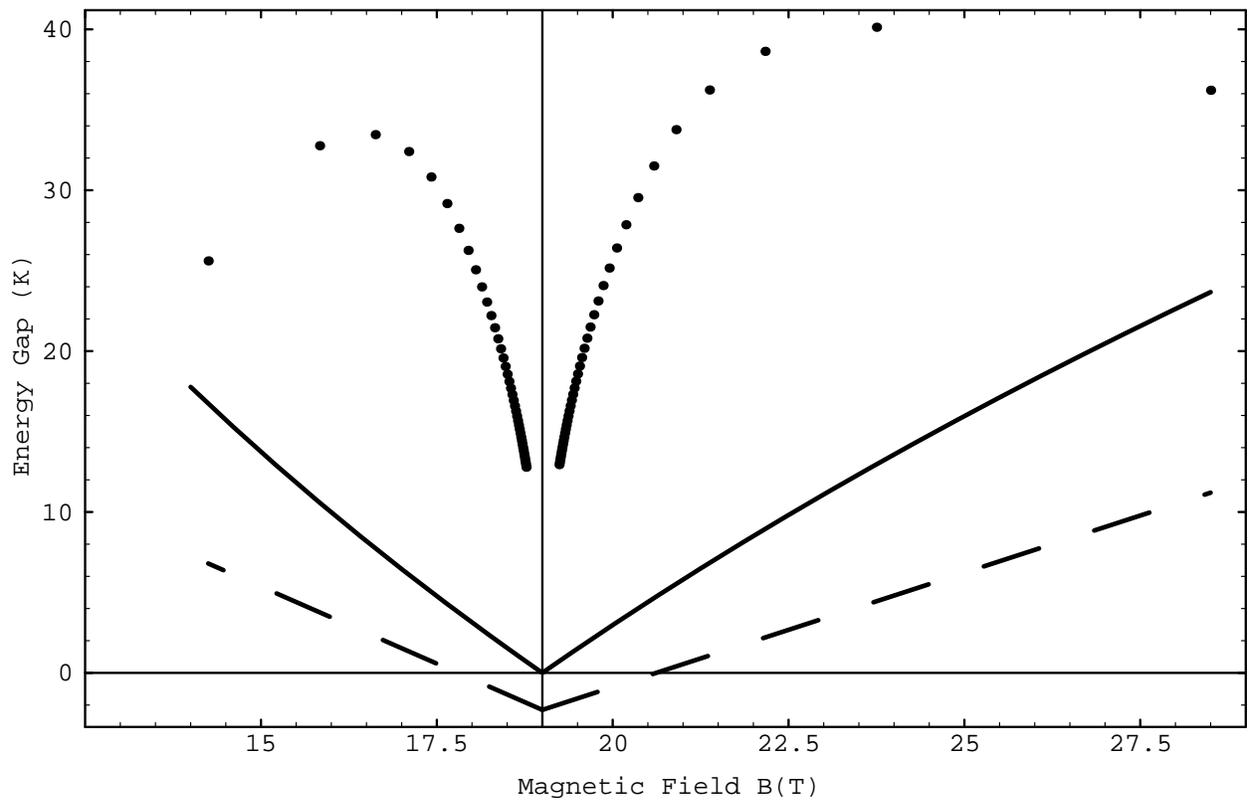
\end{document}